\documentclass[epj]{webofc}
\usepackage[varg]{txfonts}   
\wocname{EPJ Web of Conferences}
\woctitle{CONF12}
\begin{document}
\selectlanguage{english}
\title{Gauge symmetries and structure of proteins}
%
%

\author{Alexander Molochkov\inst{1}\fnsep\thanks{\email{molochkov.alexander@gmail.com}} \and
Alexander Begun\inst{1}
       \and
       Antti Niemi\inst{2}\inst{3}\inst{4}       
}
\institute{Laboratory of Physics of Living Matter, Far Eastern Federal University, 690950, Sukhanova 8, Vladivostok 
\and
           Department of Physics and Astronomy, Uppsala University, P. O. Box 803, S-75108, Uppsala, Sweden
           \and
Laboratoire de Mathematiques et Physique Theorique CNRS UMR 6083, Fédération Denis Poisson, Université de Tours, Parc de Grandmont, F37200, Tours, France
\and
Department of Physics, Beijing Institute of Technology, Haidian District, Beijing 100081, People's Republic of China
}

\abstract{%
 We discuss the gauge field theory approach to protein structure study, which allows a natural way to introduce collective degrees of freedom and nonlinear topological structures. Local symmetry of proteins and its breaking in the medium is considered, what allows to derive Abelian Higgs model of protein backbone, correct folding of which is defined by gauge symmetry breaking due hydrophobic forces. Within this model structure of protein backbone is defined by superposition of one-dimensional topological solitons (kinks), what allows to reproduce the three-dimensional structure of the protein backbone with precision up to 1A and to predict its dynamics.
}
\maketitle
\section{Introduction}
\label{intro}

Lattice gauge field theory provides universal framework for analysis of systems with infinite degrees of freedom with corresponding local and global symmetry. It allows to develop new theoretical concepts for the condensed and soft matter physics. From other other hand such systems allow direct experimental study of new nontrivial properties of the gauge field theory, what can be applied to QCD and confinement understanding. One of the most interesting and unusual subjects to which a field theory approach can be applied is the protein research.  

Currently, the most ambitious computational approaches to modelling the structure of proteins are based on classical molecular dynamics that allows us to describe the processes of protein folding in the case of short and fast folding proteins only~\cite{Feddolino}. For a realistic description of the tertiary and quaternary structures on large spatial and temporal scales computing power by 5-6 orders of magnitude greater than is technically achievable now is needed~\cite{Lindor}. 
To overcome the computational problems and to make possible the practical modelling of the properties of proteins and their complexes over large time scales, a number of models based on a rough physical assumptions has been developed. These models, increasing the speed of calculations, systematically exclude the contributions to the total atomic force fields, which are assumed to be insignificant. For example, the force field UNRES~\cite{unres} gives a very detailed, but still approximate expression for the potential energy, which includes fifteen different terms in conjunction with a simplified geometry. The considerable success of various approximate approaches allows us to raise the question - what are the factors that really important to describe the free energy? In particular, is there a systematic method to approximate the force fields, based on first principles that can reproduce the correct structure of proteins? Such a classification scheme as the CATH~\cite{cath} (http://www.cathdb.info/) and SCOP~\cite{scop} (http://scop.mrc-lmb.cam.ac.uk/scop/), who develop a taxonomic approach to describe the topology of protein available database PDB, found that despite the differences in their amino acid structures, the number of different shapes is limited. Such empirical data provide a conceptual framework for new methods for structure prediction. Within these methods folded protein structure assembled by combining fragments obtained from homologously related proteins already described in PDB. Since the number of conformations, which may be in a particular sequence is very limited, the basic idea is that various fragments, which have been described in the PDB, can be used as the designer of the building blocks for creating the folded protein. The whole structure can be obtained by coupling these fragments in the three-dimensional conformation, frequently via the energy function by comparison with similar homology by its control structures from PDB. According to the generally accepted tests Critical Assessment for Structural Prediction (http://www.predictioncenter.org/index.cgi), at the moment this type of practices have the greatest predictive power for the conformation of the protein. However, the disadvantage of these methods is the lack of sound energy function, which greatly complicates their application for research of dynamic aspects, including the processes of folding, and especially the nature of and reasons for the transition to incorrect protein folding.

  Furthermore, all of these approaches are based on data from experimental studies on synchrotron radiation sources. Currently, there is a revolutionary development of experimental techniques sources of hard gamma radiation. Now the third generation synchrotron sources such as ESFR and PETRA, next-generation sources are such facilities as the European X-ray Free Electron Laser at DESY, providing high-precision studies of the structure of proteins. However, these studies are hampered by the need for protein crystallisation. There are significant differences in the molecular structure of various proteins, and the resulting crystals are extremely fragile. Moreover, crystallisation of proteins in most cases is impossible. This leads to the fact that the number of protein structures studied by three orders of magnitude smaller than the number of known protein sequences. As a result, the gap between the number of known sequences, and studied structures remains high. In addition, the study of protein-protein interactions in the search for the causes of diseases of protein misfolding, and protein studies of conformational changes upon environment influence on receptor subunits requires much greater precision and can not be solved on the crystallised proteins. Thus, the particular relevance numerical modelling of the tertiary and quaternary structure of the protein and its dynamics in external fields in terms of collective degrees of freedom derived from first principles.

One of the solutions of the problem can be obtained with the help of coarse grain modelling within effective field theory, which allows a natural way to introduce the collective degree of freedom and nonlinear topological structures based on fundamental principles of gauge symmetry. The corresponding field theory model is based on local symmetry of proteins that will dynamically define tertiary structure of proteins~\cite{Niemi_lecture}.

\section{Local symmetry of proteins}
\label{symmetry}

Protein local symmetry is defined by amino-acids and protein backbone bonds structure. The covalent bond between amino acids central carbon atom $C_{\alpha}$ and carbon in the carboxyl group $C^\prime$ has very low rotation energy. Rotation energy of the rest of the bonds in the amino-group is high and neighbour $C_\alpha$ and all peptide bond atoms between them are in the same plane. Taking it into account we can consider the protein backbone as a freely rotating discrete chain of such planes. A such chain formally can be described by a discreet (1+1) manifold with $U(1)$ local symmetry. Correspondingly, the gauge phase can be associated with rotational angle of the $C_{\alpha}$  and $C^\prime$ bond. In the water medium the protein local $U(1)$ symmetry is broken due to hydrophobic and hydrophilic forces that lead to the amino acids alignment inside (hydrophobic) or outside (hydrophilic) of the protein secondary structure. 

The discussed symmetry breaking exhibits chirality as a consequence of the amino acids geometry. The $C_\alpha$ atom is in a $sp^3$ hybridise state, thus its covalent bonds form a regular tetrahedron with R-group, $H$, $N$ and $C^\prime$ atoms at vertices. If there is no one pair of equivalent groups at the vertices, then this structure is obviously chiral. All of the amino acids that generate a protein are chiral with left orientation except glycine, which is non-chiral due to hydrogen atom as the R-group. This symmetry breaking leads to chirality of the polypeptide ground state structures. For example, all $\alpha$-helices are right handed. 

The angles between the covalent bonds of the peptide bonds atoms have large fluctuation energy and are almost constant. Thus, the backbone bending angle fluctuates near a fixed value determined by the stable secondary structures of the protein (alpha-helix, beta-sheets and so forth). Thus, we can consider protein as a chain with almost constant bending.
   
The considered above local symmetry breaking and chirality folds the freely rotating constant bending chain into to the correct protein structure.  

The considered properties of the proteins can be formally described by an appropriate geometric description of protein. 
A such model can be formulated within the approach that use description of the local geometry of proteins based on the formalism of discrete Frenet coordinates~\cite{Niemi_lecture}. Under this formalism proteins are considered as one-dimensional discrete uniformity, which determined the free energy functional, defined solely by the angles of curvature and torsion. 
 The Frenet Frame rotation can be presented by the following transformation: 
\begin{eqnarray}
	\frac{d}{ds}\left(\begin{array}{c} {\bf e}_1 \\ {\bf e}_2\\{\bf t}\end{array} \right) = \left(\begin{array}{ccc} 0 & (\tau+\eta_s) & -\kappa cos(\eta)\\ (\tau+\eta_s) & 0 & \kappa sin(\eta) \\ \kappa cos(\eta) & -\kappa sin(\eta)& 0\end{array}\right)\times \left( \begin{array}{c}{\bf e}_1 \\ {\bf e}_2\\{\bf t} \end{array} \right)
\end{eqnarray}
 Upon rotation of the local coordinate system doublet of dynamic variables is transformed just as two-dimensional Abelian Higgs multiplet. 
This transformation can be rewritten as a gauge U(1) transformation of the scalar field:
\begin{eqnarray} 
\kappa \sim \phi \to \kappa e^{-i\eta}\equiv \phi e^{-i\eta}\\
\tau \sim A_i \to \tau + \eta_s \equiv A_i+ \eta_s	
\end{eqnarray}
Here the bending ($\kappa \sim \phi $) and torsion ($\tau \sim A_i$) are introduced as scalar and gauge filed correspondingly. 
Thus, the Abelian Higgs Model Hamiltonian takes the form:
\begin{eqnarray}
H=\int\limits_0^L ds(|(\partial_s+ie\tau)\kappa|^2+\lambda(|\kappa|^2-m^2)^2+a\tau-\frac{c}{2}\tau^2)	\label{Hmlt}
\end{eqnarray}
where $s$ is the natural parameter of the curve. The Chern-Simons term $a\tau$ is introduced to ensure chirality of proteins that leads to domination of the right hand alpha-helices, the $\tau^2$ Proca mass term follows from the partial gauge symmetry breaking. As the results we obtain ground states of the protein backbone as the theory vacua with different topological sectors: 

$\alpha$-helices (broken chiral symmetry, negative parity):
\begin{eqnarray*}
	\begin{cases} \kappa\simeq \frac{\pi}{2} \\ \tau\simeq 1 \end{cases}
\end{eqnarray*}

$\beta$-strands (restored chiral symmetry, positive parity): 
\begin{eqnarray*}
	\begin{cases} \kappa\simeq 1 \\ \tau\simeq \pi \end{cases}
\end{eqnarray*}


Loops can be considered as spatial transition between different ground states. To get explicit geometrical description of the transitions, one can apply following $Z_2$ symmetry transition of the bending and torsion: 
\begin{eqnarray}
\kappa \to -\kappa \\
\tau \to \tau+\pi.
\end{eqnarray}
Such transition leaves the geometrical structure of a one-dimensional curve unchanged and introduce inflection points in the protein chain where the curve becomes flat. These inflection points are stable topological structure that cannot be removed by a continuous transformation. Such inflection points are defined by the exact soliton solutions of the real part of the Hamiltonian (\ref{Hmlt})~\cite{Niemi2}:
 \begin{eqnarray}
 	\kappa(s)=m tanh(m\sqrt{\lambda}(s-s_0))
 \label{Soliton}\end{eqnarray}
 with energy:
 \begin{eqnarray}
 	E=\int ds (\kappa_s^2+\lambda(\kappa^2-m^2)^2-a\tau-b\kappa^2\tau+\frac{c}{2}\tau^2+\frac{d}{2}\kappa^2\tau^2)
\label{FreeEn} \end{eqnarray}
This approach allows parametrisation of the protein backbone structure by  superposition of the one-dimensional solitons (kinks) with high accuracy~\cite{Niemi2}, and dynamical properties of the protein chain is defined by the corresponding soliton dynamics.   
. 

\section{Monte-Carlo numerical simulations of the protein structure}
\label{MonteCarlo}
 To obtain the soliton structure that describes protein topology and dynamics, one have to define positions of the inflection points that fit protein tertiary structure and find soliton parameters that correspond to the global minimum of the free-energy~(\ref{FreeEn}). 
 \begin{figure}[h]
\centering
\includegraphics[width=12cm,clip]{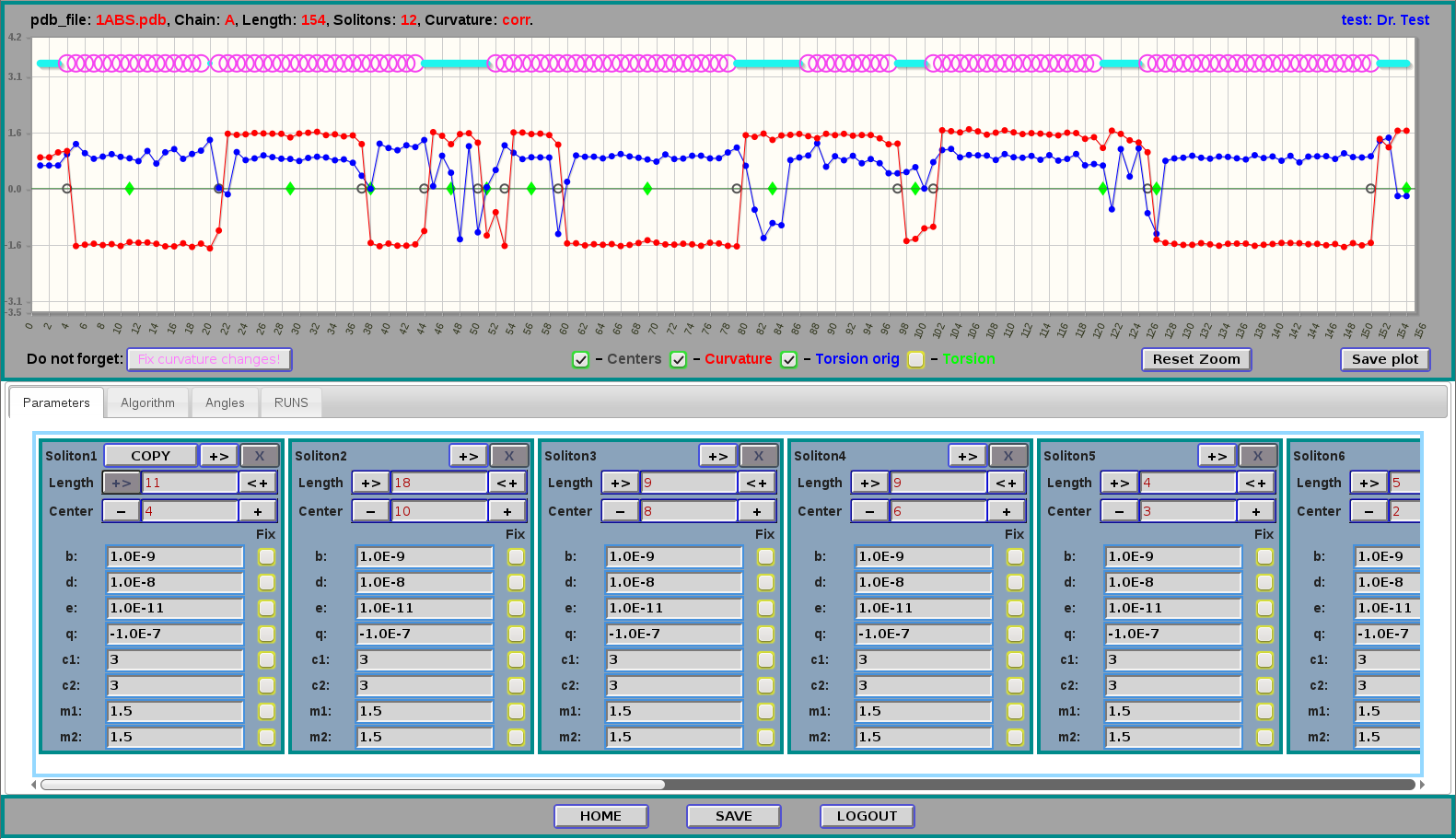}
\caption{Front-end for the protein structure visualisation and analysis toolkit. Upper panel - torsion and bending angels, lower panel - parameters for the protein soliton structure.}
\label{WebSite1}       
\end{figure}
 
The inflection point can be defined by a visual analysis of the protein topology. For this propose a special toolkit for protein structure visualisation and analysis was developed~\cite{WebSite}.  
On the figure~(\ref{WebSite1}) the front end of the toolkit is presented. In this example the torsion and bending of the Myoglobin (1ABS) is presented before and after definition of all inflection points.  
 
The free energy~(\ref{FreeEn}) exhibits two energy scales. The higher energy scale is related to the bending $\kappa$ and presented by terms that have exclusively $\kappa$ contribution. Values of the $\kappa$, that correspond to classical minimum of the free-energy, are defined by the corresponding exact solution~(\ref{Soliton}). As the first approximation we obtain parameters of the equation~(\ref{Soliton}) that fit the particular protein backbone bending with all inflection points taken into account.

\begin{figure}[h]
\centering
\includegraphics[width=11cm,clip]{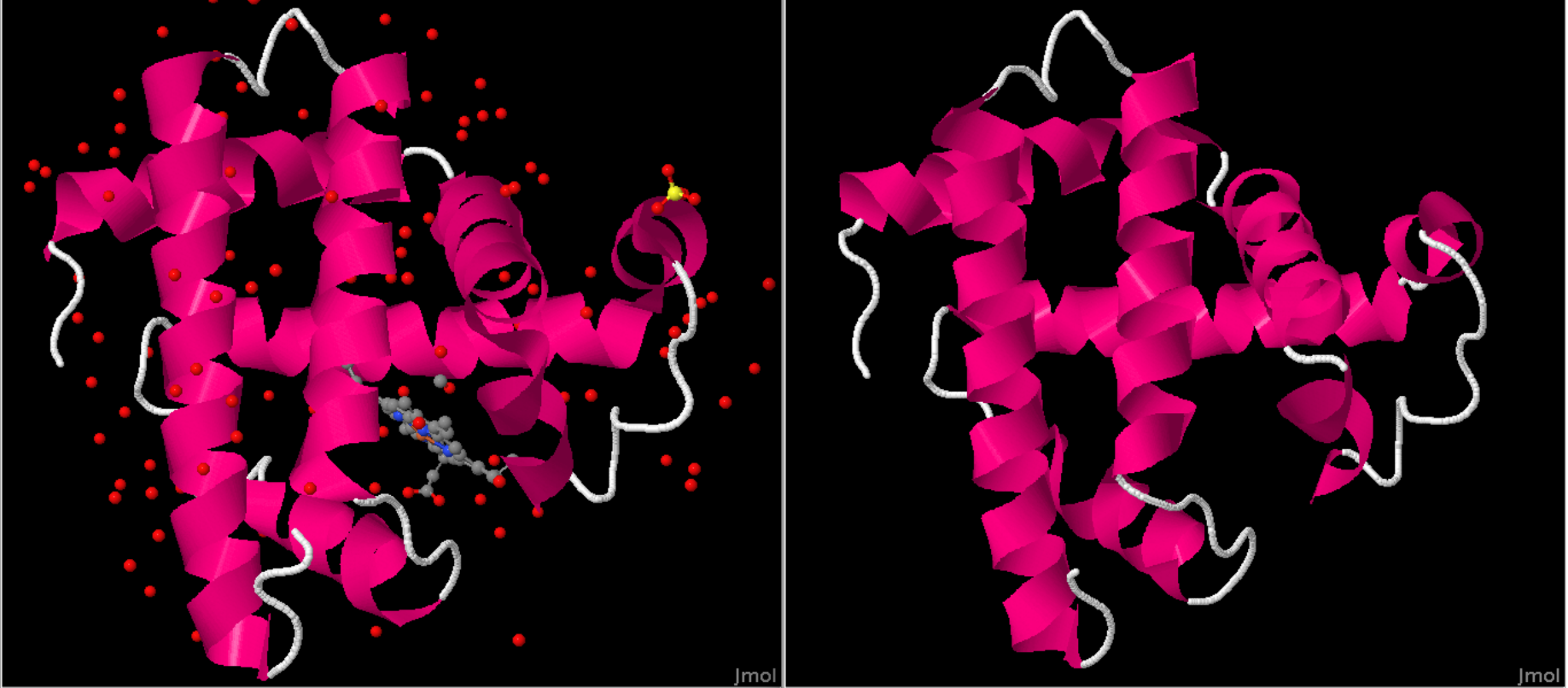}
\caption{Left panel - native three-dimensional structure of the Myoglobin from PDB. Right panel - structure reproduced by with the present approach as superpositIon of topological solitons.}
\label{Myoglobin}       
\end{figure}

The lower energy scale is related to the terms that depend off the torsion $\tau$. An extremum condition of the free-energy gives the following values of the torsion:
\begin{eqnarray} 
\tau = \frac{a+b\kappa^2}{c+d\kappa^2}	
\label{SolTrosion}\end{eqnarray}
Protein free energy can have multiple local extrema that satisfy this relation. Thus, as the next approximation we obtain parameters of the expression~(\ref{SolTrosion}) that fit the protein backbone torsion and correspond the global minimum of the free-energy, using combination of the simulated annealing and gradient descent methods.   
As the result we get precise description of the protein structure. For example we get three-dimensional structure of the Myoglobin with RMSD less then $1A$~(figure \ref{Myoglobin}.).

 This approach allows not only geometrical description of the protein three dimensional structure, but its dynamics as well. To make simulation of the protein structure thermal dynamics we apply Glauber heating algorithm~\cite{Myoglobin1,Myoglobin2}. 
 On each step of the heating algorithm we put a fluctuation of the torsion or bending at a random amino acid in a way that the new value of the torsion has the form: 
 \begin{eqnarray}
 	\tau_i^\prime = \tau_i + 1.4 R 
 \end{eqnarray}
 and new value of the bending has the following form:
 \begin{eqnarray}
 	\kappa_i^\prime = \kappa_i + 0.06 R 
 \end{eqnarray}
where $R$ - is a random number with normal distribution with the central value equal $0$ and dispersion equal $1$. 
The corresponding configuration is accepted with the probability:
\begin{eqnarray}
P=\frac{e^{-(E^\prime - E)/T}}{1+e^{-(E^\prime - E)/T}},
\label{GlaubProb}\end{eqnarray}
 where $T$ - Glauber temperature, $E$- free-energy of the protein chain before the fluctuation, $E^\prime$ - the free-energy after the fluctuation. If the configuration accepted, then the value of $\tau_i$ or $\kappa_i$ is substituted by $\tau^\prime_i$ or $\kappa^\prime_i$ correspondingly. If the configuration is not accepted then value of the angle remains same as on previous step. Then, temperature changes on a small value and on the next step the procedure is repeated. 
 \begin{figure}[h]
\centering
\includegraphics[width=10cm,clip]{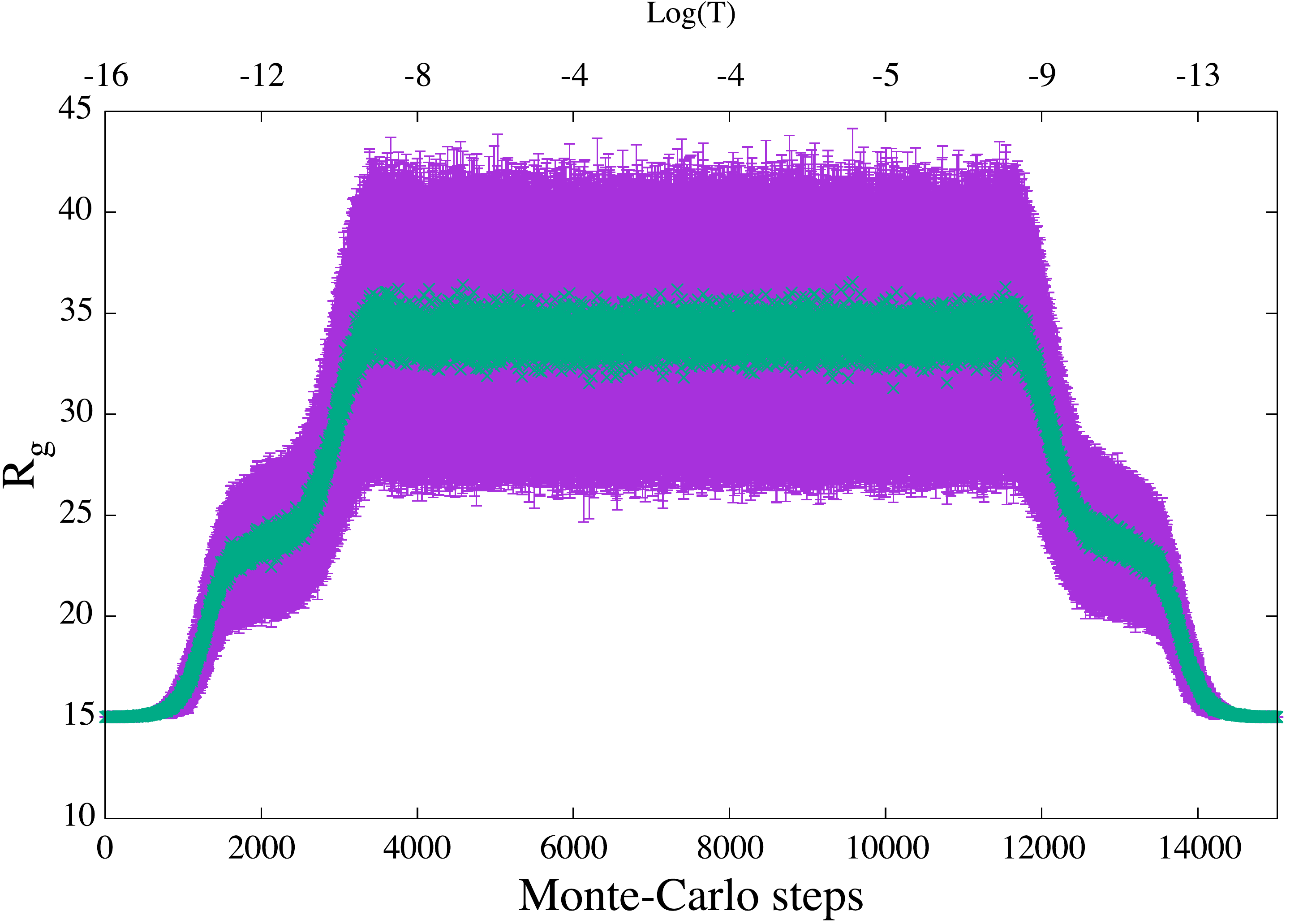}
\caption{Myoglobin radius of gyration dependence on the Glauber temperature. The x-axis - radius of gyration in Angstroms, the y-axis is the number of step in heating/cooling procedure}
\label{heating}       
\end{figure}
 \begin{figure}[h]
\centering
\includegraphics[width=10cm,clip]{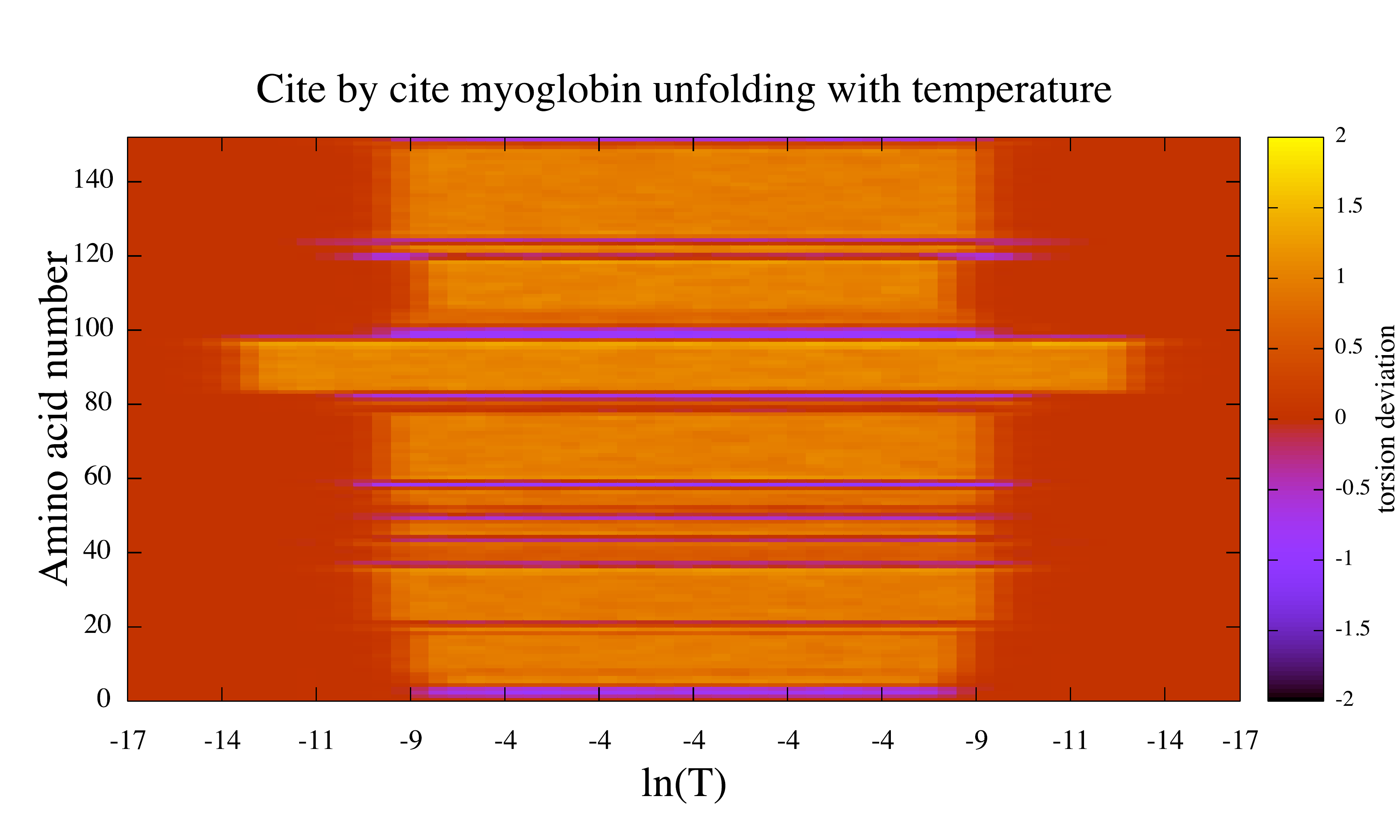}
\caption{The torsion angle deviation from the values at the Myoglobin native state (colour bar) site by site. The x-axis - logarithm of the Glauber temperature, the y-axis is the number of the amino-cid}
\label{f-helix}       
\end{figure}  

\begin{figure}[h]
\centering
\includegraphics[width=10cm,clip]{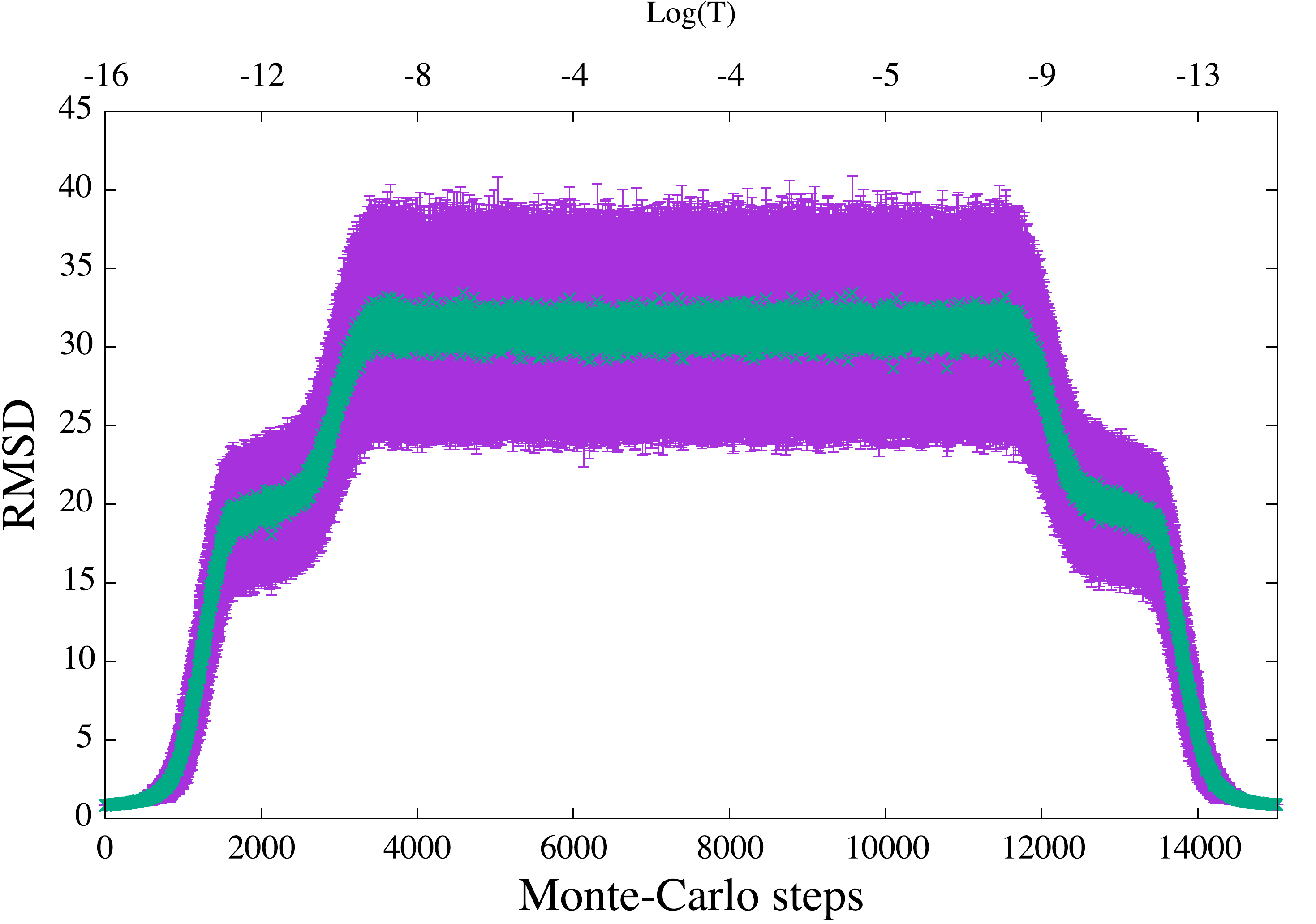}
\caption{Temperature dependence of the root-mean-square deviation of atomic positions from the native state.}
\label{RMSD}       
\end{figure}

 The result of calculation for the Myoglobin is presented at the figure~\ref{heating}.
 The simulation is performed in the Glauber temperature range from $10^{-18}$ to $10^{-4}$, with fifteen million steps.  As it follows from the expression~(\ref{GlaubProb}), the acceptance rate of the steps, were energy increases on the value larger than temperature, will be exponentially suppressed. Due to this, there is no thermal changes of the higher energy scale $\kappa$-terms of~(\ref{FreeEn}). As the result all thermal fluctuations are produced by torsion.
The radius of gyration temperature dependence shows that thermal unfolding of the obtained myoglobin soliton configuration goes through several stages. At low temperature we have native configuration, then at $T=10^{-14}$ there is a crossover to the intermediate state - molten globule. At higher temperature - $T=10^{-9}$ there is a crossover from molten globule to the self-avoiding random walk (SARW). During the cooling process, myoglobin soliton structure folds from SARW state to the native state through the same stages. Calculation of the root-mean-square deviation of atomic positions (RMSD) presented at figure~\ref{RMSD} shows that after cooling the myoglobin soliton configuration folds to the exactly same native conformation with RMSD less than 1 Angstrom. Thus, we can conclude that obtained free-energy minimum corresponds the stable native conformation of the protein. 

To understand nature of the molten globule state, we analyse torsion angle fluctuations during heating and cooling processes for each amino-acid site. The result of the numerical simulations is presented at figure~\ref{f-helix}.
The figure shows that molten globule state appears due thermal unfolding of a short part of the backbone chain. The part is located at cites from 84 to 96, what exactly corresponds to the alpha-helix F. Thus myoglobin molten globule has unfolded F-helix with the rest of the structure exactly same as at native state.

\section{Conclusions}
\label{results}
Analogy between gauge fields and protein chain Frenet frames geometry  allow us to use full power of the field theory approach to study not only protein structure in terms of the collective degrees of freedom, but also dynamics at different thermodynamical conditions and in external fields. Gauge and chiral symmetry breaking provides new concept of the protein structure study in terms of topological solitons superpositions, what allows us to reproduce protein secondary and tertiary structure with accuracy better than 1 Angstrom. Analysis of protein dynamics within field theory Monte-Carlo simulations shows that the gauge symmetry restoration at particular protein clusters provides fundamental mechanism of the protein folding to different conformations, what can be directly related to the biological function of the protein. For example, myoglobin thermal unfolding to the molten globule and folding back occurs due to unfolding and folding of the F-helix, which is bound with the heme-group. What can explain mechanism of the oxygen storage and release in myoglobin.    
{\bf Acknowledgement:}The work was supported by the Federal Target Programme
for Research and Development in Priority Areas
of Development of the Russian Scientific and Technological
Complex for 2014-2020 (The unique identifier of the Contract RFMEFI58415X0017, contract number 14.584.21.0017).


\end{document}